\documentclass[12pt,a4paper]{article}

\usepackage{graphicx}
\usepackage{wrapfig}

\usepackage{enumitem} 
\usepackage{pdfpages} 

\usepackage[T1]{fontenc} 
\usepackage{tgtermes} 

\usepackage{tabularray} 
\usepackage{graphicx} 
\usepackage[margin=25mm]{geometry}
\usepackage{amsmath}
\usepackage{amsfonts}
\usepackage{amssymb}
\usepackage{siunitx}
\usepackage{verbatim}
\usepackage{natbib} 
\usepackage{microtype} 

\usepackage[hyphens]{url}
\usepackage[hidelinks]{hyperref}
\usepackage[hyphenbreaks]{breakurl}

\title{Combatting deepfakes: \\
Policies to address national security threats and rights violations}
\author{
  Andrea Miotti$^*$\\
  \small{Control AI}\\
  \small\texttt{andrea@controlai.com}
  \and 
  Akash Wasil$^*$\\
  \small{Control AI}\\
  \small\texttt{akash@controlai.com}
}\date{}

\footnotetext{Equal contribution}

\begin{document}
\maketitle

\begin{abstract}

\noindent
This paper provides policy recommendations to address threats from deepfakes. First, we provide background information about deepfakes and review the harms they pose. We describe how deepfakes are currently used to proliferate sexual abuse material, commit fraud, manipulate voter behavior, and pose threats to national security. Second, we review previous legislative proposals designed to address deepfakes. Third, we present a comprehensive policy proposal that focuses on addressing multiple parts of the deepfake supply chain. The deepfake supply chain begins with a small number of model developers, model providers, and compute providers, and it expands to include billions of potential deepfake creators. We describe this supply chain in greater detail and describe how entities at each step of the supply chain ought to take reasonable measures to prevent the creation and proliferation of deepfakes. Finally, we address potential counterpoints of our proposal. Overall, deepfakes will present increasingly severe threats to global security and individual liberties. To address these threats, we call on policymakers to enact legislation that addresses multiple parts of the deepfake supply chain.

\end{abstract}
\newpage
\section*{Executive summary}

\textbf{Deepfakes} are non-consensually AI-generated voices, images, or videos that a reasonable person would mistake as real. Deepfakes usually involve \textbf{sexual imagery, fraud,} and \textbf{misinformation}. They also pose serious threats to \textbf{election integrity} and \textbf{national security}.

\textbf{Deepfakes pose serious threats to personal liberty and global security, and government action is needed.} Between 2022 and 2023, deepfake sexual content increased by over \textbf{400\%}, and deepfake fraud increased by \textbf{3000\%} (Home Security Heroes, 2023; Onfido, 2024). Deepfakes have also been used to impersonate public officials, spread propaganda, and commit cybercrime. With many major countries facing elections in 2024, the widespread creation and proliferation of deepfakes also present a growing threat to democratic processes around the world. 

\textbf{The only effective way to stop deepfakes is to hold entities across the supply chain accountable for deepfake creation and proliferation.} The deepfake supply chain starts small and ends up large: a few groups supply the technology used to make deepfakes, whereas billions of people worldwide can access the technology to create a deepfake. To effectively address deepfakes, all parties in the supply chain must show that they have taken reasonable steps to preclude deepfakes. This approach is similar to how society addresses child sexual abuse material and malware. 

\textbf{Governments must ban deepfakes across the supply chain.} Effective legislation would:
\begin{enumerate}
    \item \textbf{Make the creation and dissemination of deepfakes a crime} and allow people harmed by deepfakes to sue for damages.  
    \item \textbf{Hold model developers liable for negligence.} They must show that they have applied techniques to prevent their models from being used to create deepfakes. This includes: 1) precluding a model’s ability to generate deepfake pornography or commit fraud; 2) showing that such techniques cannot be easily circumvented, and; 3) guaranteeing that the datasets they use to train their model do not contain illegal material (e.g., child sexual abuse material).   
    \item \textbf{Hold model providers and compute providers liable for negligence.} They must show that they have applied techniques to prevent their resources from being used to create deepfakes. This includes: 1) implementing reasonable measures to monitor users; 2) detecting users that are trying to create deepfakes, and; 3) restricting access from malicious users.

\end{enumerate}

\textbf{These proposals are necessary to address the growing threats to global security and personal liberty posed by deepfakes.} The proposals can inform efforts by legislative bodies, regulatory agencies, and standards-setting groups that aim to reduce threats from deepfakes. 


\newpage
\section{Introduction}\label{sec:intro} 

\hfill\begin{minipage}{\dimexpr\textwidth-1cm}
“Making a sophisticated fake with specialized software previously could take a professional days to weeks to construct, but now, these fakes can be produced in a fraction of the time with limited or no technical expertise. This is largely due to advances in computational power and deep learning, which make it not only easier to create fake multimedia, but also less expensive to mass produce.” – Department of Defense, 2023
\end{minipage}
\hspace{1cm}

\textbf{Deepfakes are a serious and growing threat to individuals, institutions, and governments. }Deepfakes are commonly used to generate non-consensual pornography, including sexual content of children. 99\% of the victims targeted by deepfake sexual content are women, and 99.6\% of explicit images of children are of females (Home Security Heroes, 2023; Internet Watch Foundation, 2023). Victims of deepfake pornography and sexual abuse often describe feeling sexually objectified, losing control over how their bodies are portrayed, and experiencing extreme levels of distress (Law Commission, 2021). 79\% of business leaders believe that deepfakes are a threat to their business, and over 1 in 3 businesses have experienced deepfake fraud (Regula, 2023). In addition to the clear dangers to individuals and businesses, many experts are concerned about the potential impacts of deepfakes on government institutions. Deepfakes can be used to create falsified videos of elected officials, sway the results of elections, blackmail political candidates, or spread propaganda during international conflicts. For example, a deepfake of Ukrainian President Volodymyr Zelensky appeared to direct Ukrainian soldiers to surrender to Russian forces (Congressional Research Service, 2023). The Department of Homeland Security has warned that deepfakes pose major threats to national security, and the Department of Defense has described risks from fraud, phishing, and other forms of cybercrime (Department of Homeland Security, 2021; Department of Defense, 2023).

\textbf{Effective regulation must address multiple parts of the deepfake supply chain.} Deepfakes can be created by anyone around the world. Once the software capable of generating deepfakes is downloaded, it cannot be withdrawn. Therefore, effective and enforceable deepfake policies would need to not only punish creators of deepfakes but also hold companies liable if they fail to take reasonable steps to prevent deepfakes. This includes provisions for model developers (e.g., applying techniques to prevent models from generating deepfakes) as well as model providers and compute providers (e.g., applying techniques to identify users that are trying to create deepfakes and restricting access from malicious users). This approach is also consistent with the idea of “systemic regulation” in AI, which has highlighted the need to regulate AI as a technology as opposed to focusing solely on its downstream applications (see Arbel et al., 2023).

\textbf{The primary goal of this report is to outline concrete policy proposals that governments can use to effectively ban deepfakes.} In section 2, we review existing proposals to address threats from deepfakes. In section 3, we detail our main policy recommendation: a ban on deepfakes that covers multiple parts of the deepfake supply chain.

\newpage
\section{Review of previous proposals to address deepfakes}

\hfill\begin{minipage}{\dimexpr\textwidth-1cm}
“I urge you to consider legislation and regulation to address the misuse of deepfake technology as a whole… To address the challenges posed by deepfakes, tech companies should be responsible for developing and implementing the policies to detect and mitigate this type of content.” – Professor David Doermann testifying before the US House Committee on Oversight and Accountability (Doermann, 2023)
\end{minipage}
\hspace{1cm}

There have been several legislative proposals designed to address risks from deepfakes. For example, The Preventing Deepfakes of Intimate Images Act establishes criminal penalties for individuals who share deepfakes of intimate images, and the DEEPFAKES Accountability Act provides criminal penalties for individuals who create malicious deepfakes without labeling them as deepfakes. The NO FAKES Act and the NO AI FRAUD Act go further, making companies liable for contributing to the creation or spread of deepfakes. This approach is essential given the nature of the deepfake supply chain (as we describe in the next section). 

Below, we offer more details about existing legislative proposals:

\textbf{Preventing Deepfakes of Intimate Images Act (HR 3106).} This bill would prohibit the non-consensual disclosure of digitally-altered intimate images and make the sharing of these deepfakes a criminal offense. It would also create a right of private action for victims of such deepfakes and provide protections to allow plaintiffs to preserve their anonymity. The bill was introduced in May of 2023 with 20 co-sponsors and was referred to the House Committee on the Judiciary (H.R.3106 - 118th Congress, 2023-2024). 

\textbf{DEEPFAKES Accountability Act (HR 5586).} This bill would establish criminal penalties for individuals who create malicious deepfakes and fail to label them as deepfakes. The term “malicious deepfakes” refers to deepfakes relating to sexual content, violence or physical harm, financial fraud, or influencing public policy debates and elections. If an individual creates a malicious deepfake with the intent to distribute it, and they fail to label it as a deepfake, they could face criminal or civil penalties under this bill. The bill was introduced in September of 2023 with one co-sponsor and was referred to three committees: the House Judiciary Committee, the Energy and Commerce Committee, and the Homeland Security Committee (H.R.5586 - 118th Congress, 2023-2024). 

\textbf{NO FAKES Act (draft bill, Senate).} This draft bill would grant people the right to authorize their voice or visual likeness. The bill would hold individuals and companies liable if they create or distribute an unauthorized digital replica of an individual, and it would hold platforms liable for hosting unauthorized deepfakes. A draft version of the bill was released in October 2023 by four Senators (Senate Legislative Counsel, 2023). 

\textbf{NO AI FRAUD Act (HR 6943).} Like the NO FAKES Act, this draft bill would also grant people the right to authorize their voice or visual likeness. The bill would hold entities liable if they “materially contribute to, direct, or otherwise facilitate” the distribution or publication of deepfakes. The bill also has a clause about First Amendment defenses, providing criteria that courts can use to balance First Amendment protections against intellectual property rights. A draft version of the bill was released in January 2024 by two Representatives (NO AI FRAUD Act [Draft Bill], 2024). The bill was introduced in January of 2024 with 7 co-sponsors and was referred to the House Judiciary Committee (H.R.6943 - 118th Congress, 2023-2024). 

\textbf{Online Safety Act 2023 (UK).} The Online Safety Act is a broad piece of legislation that covers protections for children, cyberbullying and harassment, data privacy and security, and other areas of internet safety. One section of the Online Safety Act outlaws the non-consensual sharing of deepfake intimate images (Ministry of Justice, 2022). Importantly, the Online Safety Act places legal obligations on companies to identify, mitigate, and manage risks of harm from illegal content. Social media companies are required to proactively identify and mitigate illegal content, especially content that could be harmful to children. Companies that fail to comply face major financial penalties (up to 10\% of global annual revenue), and senior employees could face criminal charges for non-compliance. The Act’s approach to regulating deepfakes followed the recommendations of a Law Commission report (see Law Commission, 2022). Although the Act provides a strong start by placing liability on social media companies, it lacks regulations for AI model developers or model providers. This means that deepfakes are often shared millions of times on social media before they are detected or removed. To extend the work of the Online Safety Act, we need not only to prohibit deepfake proliferation (via regulations on social media companies) but also deepfake creation (via regulations across the entire supply chain, including model developers and model providers). The Online Safety Act became law in October of 2023 (Online Safety Act 2023). 

\textbf{Criminal Justice Bill of 2023-2024 (UK).} The Criminal Justice Bill is a broad piece of legislation that aims to reform the UK’s criminal justice system. Some sections of the Criminal Justice Bill address the creation and proliferation of non-consensual intimate images, including deepfakes (Criminal Justice Bill, 2023a). Importantly, the Criminal Justice Bill includes liability for entities that create equipment with the intention of enabling the creation of intimate images. As a result, entities that develop or provide AI models, tools, and guides for deepfake creation could be liable. However, prosecutors may need to prove “intent”, arguing that the AI model or tool is explicitly designed with the intention to enable the creation of non-consensual intimate images. To avoid liability, it is possible that entities could simply claim that they do not intend to proliferate non-consensual deepfakes, even if they provide models and guides specifically designed to make intimate deepfakes. For example, they could claim that they are merely intending to help people engage in consensual image manipulation, even if the vast majority of their users are engaging in non-consensual deepfakes. The Criminal Justice Bill was introduced in the House of Commons in November 2023 (Criminal Justice Bill, 2023b). 

\newpage
\section{Banning deepfakes: Policy recommendations}

\hfill\begin{minipage}{\dimexpr\textwidth-1cm}
“Accountability should be distributed across technology actors involved in the production of AI technologies more broadly, including from the foundation models to those designing and deploying software and apps, and to platforms that disseminate content.” – Sam Gregory testifying before the US House Committee on Oversight and Accountability (Gregory, 2023).
\end{minipage}
\hspace{1cm}

\textbf{Governments should ban deepfakes by establishing liability across the supply chain.} The deepfake supply chain starts small (a few groups supply the technology used to make deepfakes) and ends up large (billions of people around the world can access the technology to create a deepfake). Given the nature of the deepfake supply chain, deepfakes will proliferate unless there are regulations across the supply chain. Moreover, policy cannot proceed on the assumption that deepfakes will be identified as deepfakes. We cannot only make deepfakes illegal: we also need to make it extremely difficult to create them. This is the only effective way to stop deepfakes: if regulations only target users, deepfakes will continue to spread. 

Below, we describe the deepfake supply chain, and we propose regulations for entities across the supply chain. 


\subsection{Understanding the deepfake supply chain}

\textbf{The deepfake supply chain consists of several steps.} The supply chain consists of model developers, model providers, compute providers, and deepfake creators. Put simply, for a deepfake to be created, a \textbf{model developer} needs to train an AI model. Then, a \textbf{model provider} needs to provide the AI model to users. In some cases, model providers offer “tweaked” versions of models that have been edited to make them better at creating deepfakes. To develop a model or use a model, computational resources are required.  

\textbf{Compute providers} offer computational resources to model developers (to build AI systems) and to users (to run AI systems). Finally, \textbf{deepfake creators} can create deepfakes by accessing models (from model providers) and computational resources (from compute providers, or using local hardware). 

Figure 1 illustrates the deepfake supply chain.

\begin{figure}[ht]
\centering
\includegraphics[width=0.47\linewidth]{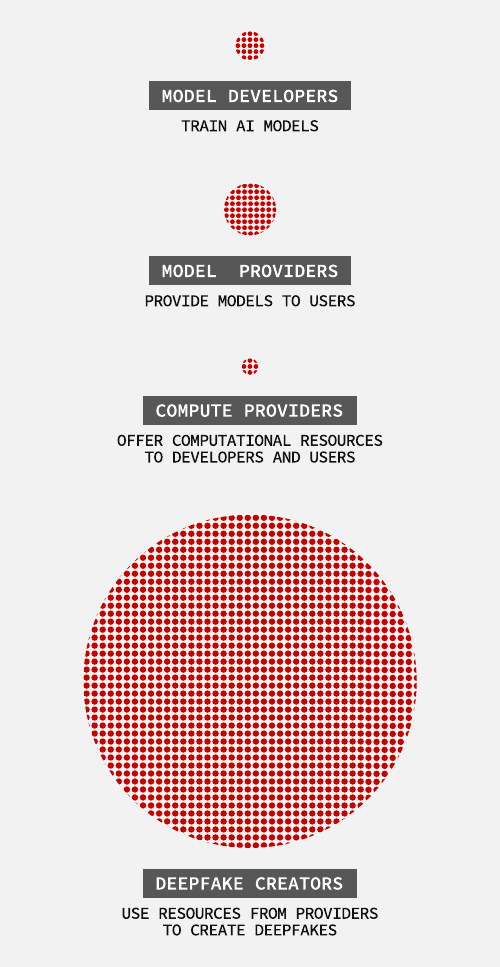} 
\caption{Representation of the deepfake supply chain. The supply chain begins narrow with a relatively small number of model developers, model providers, and compute providers. At the end of the supply chain, an extremely large number of users have the potential to create deepfakes.}
\label{fig:myLabel}
\end{figure}

\textbf{Model developers} refer to a handful of groups that can build powerful generative AI systems. Developing these models generally requires large amounts of computational resources and strong technical talent. Currently, only a few major groups have developed such models. Examples include OpenAI (DALL-E), Google (Imagen), StabilityAI (Stable Diffusion), and Midjourney. A few groups have developed models that are specifically tailored for producing high-quality deepfakes; examples include DeepFaceLab and FaceSwap.

\textbf{Model providers} refer to entities that offer models to users. Many model developers are also model providers (for example, OpenAI developed DALL-E and also provides DALL-E to users). In addition, some entities do not develop their own models but offer models to users. As an example, DeepFaceLab offers software, tools, and models specifically for creating deepfakes. MrDeepFakes forum – one of the most popular online forums for discussing deepfakes for non-consensual pornography and sexual abuse material – offers models in addition to hosting deepfakes and allowing users to request deepfakes of specific individuals (Timmerman et al., 2023). Github, a platform that offers code and software for various types of AI development, currently hosts code and software that is designed specifically for deepfakes (including code and software made by DeepFaceLab and users of the MrDeepFakes forum). 

Importantly, there are two different ways to provide models, and these different methods have major implications for deepfake creation. Some model providers provide their models behind a web application or interface. This gives the model provider more control over how users engage with their AI system: model providers have the ability to detect malicious use, withdraw access from specific users, and even withdraw access from all users if a model needs to be tweaked. OpenAI, for example, offers access to DALL-E through an Application Programming Interface (API): users need to make an account and obtain an API key from OpenAI. This allows OpenAI to implement several measures to reduce malicious use. Examples include controlling which users have access to the API, setting limits on the number of requests a user can make in a certain time frame, monitoring and logging usage to detect patterns of malicious use, filtering certain kinds of requests, enforcing terms of service agreements, updating or patching the model if safety issues are discovered, and conducting audits to ensure that API usage complies with legal requirements. In contrast, DeepFaceLab offers image-generation tools without an API. Users can download the tools directly onto their computers. DeepFaceLab cannot monitor use, restrict access from malicious users, or update the software that users downloaded. This makes DeepFaceLab’s tools much more appealing from the perspective of deepfake creators: there is no oversight and no ability for DeepFaceLab to stop the proliferation of deepfakes. 

\textbf{Compute providers} refer to companies that offer computational resources to build or run powerful AI systems. Developing an AI system requires significantly more computational resources than running an AI system. When someone runs an AI system, computational resources are often provided by the model provider, but some individuals and groups purchase their own computational resources. For large-scale operations – like training an advanced AI system or providing access to large numbers of users – computational resources are provided by a small number of large companies. Examples include Amazon (Amazon Web Services), Microsoft (Microsoft Azure), Google (Google Cloud Platform), and IBM (IBM Cloud). 

\textbf{Deepfake creators} refer to individuals who access models and computational resources to create deepfakes. If the technology and the computational resources are available, creating a deepfake is relatively simple. Creating deepfakes does not require AI knowledge or programming skills. Some websites offer apps that allow anyone to freely create deepfakes in less than 30 seconds (Finger, 2022). As a result, there are billions of potential deepfake creators, and deepfake creation is expected to become even easier as technology advances.

\subsection{Regulations at each step of the supply chain}

\textbf{To ban deepfakes, governments should establish liability across the supply chain.} In this section, we describe regulations for deepfake creators, model developers, model providers, and compute providers. When relevant, we cite relevant sections of previous proposals or regulations in other sectors.

\textbf{Model developers should be liable for negligence if they fail to take reasonable steps to prevent models from creating deepfakes.} Model developers have a responsibility to design models using techniques that prevent them from creating deepfakes. Such measures could include techniques that reduce a model’s ability to generate deepfake sexual material or fraudulent content, implement techniques that cause models to refuse requests to generate deepfake sexual material or fraudulent content, and show that such techniques cannot be easily circumvented. A relevant precedent can be found in hardware manufacturers: manufacturers must ensure GPS-enabling chips no longer function when used above certain speeds (to prevent their dual-use in weaponry). 

Additionally, model developers should have to guarantee that the datasets they use to train their model do not contain illegal material (e.g., child sexual abuse material). Recently, a team of researchers discovered child sexual abuse material in LAION-5B, a large image dataset that was used to train many image models – including the popular Stable Diffusion model (Stanford Internet Observatory, 2023). In response, the organization responsible for LAION-5B withdrew its dataset. This decision was entirely voluntary; it ought to be incorporated into law, and model developers should be liable unless they use reasonable techniques to ensure that their models are not trained on datasets with illegal content.  

\textbf{Model providers and compute providers should be liable for negligence if they fail to take reasonable steps to monitor how their resources are used and prevent users from creating deepfakes.} Model providers and compute providers have a responsibility to ensure that malicious actors are detected and prevented from creating deepfakes. Such measures could include techniques to detect users who are trying to create deepfake sexual material or fraudulent content, ensuring that model access is provided via an Application Programming Interface (API) whose access can be withdrawn, requiring that users register with a verified account, and other techniques designed to detect malicious users and restrict their access. A relevant precedent can be found in banking: banks are required to monitor and prevent customers from engaging in money. To prevent money laundering, financial institutions have to follow “Know Your Customer” (KYC) regulations that require them to verify the identity of their customers, assess the risk that customers may be involved in illegal activities, monitor client transactions, and report illegal activities to relevant authorities. KYC requirements for compute providers have been recommended previously in multiple contexts. Microsoft President Brad Smith testified to Congress that KYC provisions could help prevent theft, fraud, and misuse of AI systems (Smith, 2023). Furthermore, AI experts have argued that KYC requirements can help governments identify and mitigate sudden threats from AI systems (Egan \& Heim, 2023). To illustrate the importance of provisions on compute providers, the Civitai case provides a useful real-world example. Civitai developed a dataset that included child sexual material. Civitai was receiving computational resources from Amazon Web Services, but Amazon was not required to detect misuse or respond. After the child sexual material in Civitai’s dataset was discovered (by an independent team of researchers unaffiliated with Amazon), compute resources were withdrawn (Multiplatform.AI, 2023). The decision to withdraw compute was entirely voluntary, and Amazon was not required to take reasonable measures to detect such misuse. In this case, it was fortunate that the illegal material was identified and compute access was voluntarily withdrawn. However, many such cases will go undetected and unaddressed unless compute providers and model providers are held legally accountable.

\textbf{The creation and dissemination of deepfakes should be a crime.} Individuals who create or spread deepfakes should be criminally and civilly liable. Legal experts have argued that protecting victims from deepfakes may require criminal penalties (e.g., Law Commission, 2022). This principle is present in previous proposals, like the Preventing Deepfakes of Intimate Images Act (which makes the non-consensual sharing of intimate deepfakes a crime) and the DEEPFAKES Accountability Act (which establishes criminal penalties for individuals who create malicious deepfakes without labeling them as such). These measures expand on existing rules that outlaw the creation of child sexual abuse material (CSAM) that is not AI-generated. 

\subsection{Example scenarios}

In this section, we provide a few illustrative examples of deepfakes in society, briefly describe how existing laws handle these deepfakes, and describe how our proposal would change the status quo. 

\textbf{Case: Biden deepfake in New Hampshire primary.}

\textbf{Description:} In January 2024, a deepfake of Joe Biden interfered with the New Hampshire primary. Potential voters in the New Hampshire primary received a call that appeared to come from Joe Biden. The fake Biden voice said, “it’s important that you save your vote for the November election… voting this Tuesday only enables Republicans in their quest to elect Donald Trump again” (NBC News, 2024b).  
\textbf{Existing laws:} Under existing laws, the creator of this deepfake (if identified) might be pursued under federal election fraud laws. However, there would be no consequences for any other members of the supply chain: those providing the services required to make this deepfake will not have to change anything. Malicious users could download deepfake software on their personal computers and avoid any degree of oversight. As a result, we would continue to see deepfakes used for election-related fraud (likely at an increasing rate until the election occurs). Thus, existing laws would not be sufficient to prevent malicious actors from continuing to create deepfakes to tamper with election outcomes.

\textbf{Our proposal:} In contrast, under our proposed policy, the creator of this deepfake would be unambiguously civilly and criminally liable. If it were uncovered that the Biden deepfake was created using services from a negligent entity (e.g., a model provider that failed to engage in appropriate monitoring or verification), then that company would be sued and face appropriate penalties. As a result, companies would have an incentive to prevent such deepfakes from occurring. Malicious users would only be able to access deepfake technology through APIs, and companies would have measures in place to prospectively and retroactively identify malicious model requests. Thus, our proposal would substantially reduce the ability of malicious actors to create election-related deepfakes.

\textbf{Case: Deepfakes of Sunak promote investment scam.}

\textbf{Description:} In January 2024, over 100 deepfakes of Rishi Sunak circulated on social media websites. Over \$15,000 was spent to promote the deepfakes on Facebook, and estimates suggest the deepfakes may have reached 400,000 people. In one of the deepfakes, BBC news reports that Sunak was caught “earning colossal sums from a project that was initially intended for ordinary citizens.” The deepfake then encourages users to invest in a new investment platform allegedly created by Elon Musk (Fenimore Harper, 2024). 

\textbf{Existing laws:} To our knowledge, no legal action was taken against the group that produced the deepfakes. In fact, just months earlier, the group responsible for these deepfakes had also been caught in a separate deepfake investment scam. 

\textbf{Our proposal:} Under our proposed policy, creating deepfakes for financial fraud would be outlawed. Furthermore, AI model providers would have a responsibility to withdraw model access from groups that use AI to conduct such scams. As a result, fewer scams would occur in the first place, and those that do would be unambiguously illegal. 

\textbf{Case: Deepfake sexual abuse of high school students in New Jersey.}

\textbf{Description:} In October 2023, over 30 female high school students in New Jersey learned that they had been non-consensually depicted in sexual deepfakes. The deepfakes were created by male classmates (NBC News, 2024a). Reports suggest that the victims do not know the identities of the individual(s) who made the deepfakes, if the deepfakes still exist, and what punishments (if any) the school district implemented (NBC News, 2023). 

\textbf{Existing laws:} A New Jersey State Senator reported that the incident would rise to a “cyber-type harassment claim”, even though he believed that it should be prosecuted as a more serious crime (NBC News, 2023). Federal child sexual abuse material (CSAM) laws outlaw the creation and possession of child pornography, but it is unclear if they will be applied to this case. Additionally, some states have laws specifically against deepfake sexual abuse material, but most do not. As of now, New Jersey does not have any laws penalizing individuals for the creation or disclosure of deepfakes, though a state bill has been introduced (N.J. Legis., 2023). Overall, the case has called attention to the lack of clear and standardized protections against sexual deepfakes. Absent any new countermeasures, we should expect to see many more examples of sexual deepfakes targeted primarily at women and girls. In many cases, the victims of these deepfakes will never even know that they have been deepfaked and the perpetrators will not be identified.

\textbf{Our proposal:} In contrast, under our proposal, there would be clear civil and criminal penalties for individuals creating or distributing sexual deepfakes. Furthermore, we would not rely on deepfake perpetrators being identified by victims, school districts, or local law enforcement professionals – we would instead intervene earlier in the supply chain. Entities providing technology capable of producing deepfakes would share the responsibility of detecting and preventing deepfakes. We would not only make punishments clearer when deepfake perpetrators are caught – we would make it harder for them to create deepfakes in the first place and harder for them to evade detection. 

\hfill\begin{minipage}{\dimexpr\textwidth-1cm}
“No kid, teen or woman should ever have to experience what I went through. I felt sad and helpless.” – Francesca Mani, victim of sexually explicit deepfakes, aged 15. (NBC News, 2024a)
\end{minipage}
\hspace{1cm}

\subsection{Counterpoints and responses}

\textbf{First Amendment protections.} Any legislation on deepfakes in the United States will have to be consistent with First Amendment protections. Relevant precedents can be found in copyright law and defamation law. Using copyrighted materials and making untrue claims that damage someone’s reputation (defamation) are two clear examples of speech that is \textit{not} protected by law. In both cases, legislators and courts have found ways to balance these limitations with First Amendment protections. In copyright law, 17 U.S.C. § 107 (from the Copyright Act of 1976) describes a fair use principle that establishes exceptions to copyrighted material. For example, work that is used for education, research, journalism, parody, and satire is often excluded from copyright protections. In contested cases, courts are instructed to examine the purpose of the use (favoring defendants that use work for non-commercial or educational purposes) and the nature of the copyrighted work (favoring defendants that use factual work like news reports, rather than creative work like novels or music). In defamation law, Supreme Court cases have established relevant exceptions. For example, in \textit{New York Times Co v. Sullivan} (1964), the Supreme Court ruled that First Amendment considerations can limit the ability of public officials to sue for defamation. Specifically, the court adopted an “actual malice” standard – in addition to proving the standard elements of defamation, public officials must also show that the defendant \textit{knowingly} made a false statement or had \textit{reckless disregard} for whether the statement was false. As another example, in \textit{Hustler Magazine, Inc v. Falwell} (1988), the Supreme Court extended First Amendment protections to cases of parody and satire. Overall, these precedents from copyright law and defamation law show that certain forms of speech can be limited in ways that are consistent with First Amendment protections. 

Deepfake laws could adhere to such precedents by establishing exceptions for certain kinds of innocuous image and voice manipulation. For example, the NO AI FRAUD Act describes possible exceptions. It notes that “alleged harms [from deepfakes] shall be weighted against: (a) whether the individual whose voice or likeness is at issue is necessary for and relevant to the primary expressive purpose of the work in which the use appears; (b) whether the use is transformative; and (c) whether the use constitutes constitutionally protected commentary on a matter of public concerns” (NO AI FRAUD Act [Draft Bill], 2024). The NO AI FRAUD Act also includes a section about First Amendment defenses, in which it states: “in evaluating any such defense, the public interest in access to the use shall be balanced against the intellectual property interest in the voice of likeness.” 

\textbf{Watermarking.} Watermarking involves embedding a statistical signal into AI-generated material, such that the material can be detected as AI-generated. Ideally, watermarking would be able to ensure that all AI-generated content can be detected, thus allowing society to clearly distinguish between AI-generated content and content that is not generated by AI. At first glance, watermarking appears to be a promising intervention. However, it suffers from two key problems. First, many deepfakes can cause harm even if they are labeled as AI-generated. Many websites that distribute deepfake sexual images make it clear that they are AI-generated, but these deepfakes still cause harm to those depicted. As an example, even if a non-consensual pornographic video of a woman starts with a disclaimer noting that the video was AI-generated, she may still experience feelings of sexual objectification, shame, and high levels of distress. Second, recent research from a team at Harvard University has shown that robust watermarking is impossible (Zhang et al., 2023a). The researchers were able to show that watermarks can be trivially eliminated, noting: 

\hfill\begin{minipage}{\dimexpr\textwidth-1cm}
“The bottom line is that watermarking schemes will likely not be able to resist attacks from a determined adversary. Moreover, we expect that this balance will only shift in the attacker’s favor as model capabilities increase. Future regulations should be based on realistic assessments of what watermarking schemes are and are not able to achieve.” – Zhang et al., 2023b
\end{minipage}
\hspace{1cm}

Robust watermarking is not possible, and even if it were possible, it would not address many of the harms from deepfakes. There are many technical challenges of watermarking that currently limit its viability (e.g., Saberi et al., 2023; Zhang et al., 2023a). As such, watermarking should not be considered a solution to problems caused by deepfakes. Watermarking can be incorporated alongside other interventions to address deepfakes, but they should not be seen as a substitute for meaningful regulation.  

\textbf{Existing laws.} In the US, there are no federal laws specifically tackling deepfakes. Additionally, existing laws against intimate image abuse and fraud are often insufficient. In the UK, the Online Safety Act 2023 places obligations on social media companies to remove illegal content, including sexual deepfakes. This is an excellent step in the right direction, as it holds a crucial part of the supply chain (social media companies) responsible. Our proposal would expand this principle by placing obligations on AI model developers and model providers. Even if social media companies remove deepfakes, they can still be shared hundreds of thousands of times before they are detected. Our proposal would ensure that deepfake proliferation is limited even earlier in the supply chain – before they are created. 

\newpage
\section{Conclusion}
Deepfakes violate individual liberties and present serious security threats. These threats are only increasing in scale and magnitude as the technology required to develop deepfakes becomes more widespread, more convincing, and easier to use. The only effective solution involves implementing measures across the supply chain. To operationalize our proposals in greater detail, our team has prepared draft legislative text for the US and the UK (interested readers are encouraged to reach out to the authors for more information). 

Deepfakes represent one of the first clear instances in which rapid improvements in AI capabilities result in threats to individuals and society as a whole. To ensure responsible and beneficial AI progress, a broad array of policy proposals will be needed (for examples, see Miotti \& Wasil, 2023a; Miotti \& Wasil, 2023b). Deepfake policy represents a clear and tangible domain in which harms from AI urgently need to be addressed. Moving forward, we expect to continue to see capabilities from AI systems that pose concrete harm both to individuals and to society at large. Policymakers have a time-sensitive opportunity to reduce threats from deepfakes while also putting in place guardrails and processes that can protect society from ever-evolving threats from AI. 

In summary, we outlined a set of proposals that policymakers can implement to address threats from deepfakes. Our proposals (and corresponding legislative text) are designed to protect consensual and innocuous applications of AI-enabled image and voice manipulation while combatting the negative consequences of deepfakes. These proposals can inform future efforts by legislative bodies and standards-setting organizations to mitigate the impact of deepfakes in domains such as non-consensual pornography, financial fraud, and election interference.

\pagebreak

\bibliographystyle{apalike}
\bibliography{example}
\sloppy

Arbel, Y, Tokson, M \& Lin, A. (2023) \textit{Systemic Regulation of Artificial Intelligence.} \url{https://ssrn.com/abstract=4666854}

Communications Act 2003, c. 21. \url{https://www.legislation.gov.uk/ukpga/2003/21/contents}

Congressional Research Service. \textit{Deep Fakes and National Security} (IF11333; Apr. 17, 2023). Prepared by Sayler, K. M., \& Harris, L. A. \url{https://crsreports.congress.gov/product/pdf/IF/IF11333}

Copyright Act. (1976). 17 U.S.C. § 107. \url{https://www.copyright.gov/title17/title17.pdf}

Copyright, Designs and Patents Act 1988, c. 48. \url{https://www.legislation.gov.uk/ukpga/1988/48/contents}

Criminal Justice Bill. (2023a, November 14). Parliament: House of Commons. Bill 10. \url{https://publications.parliament.uk/pa/bills/cbill/58-04/0010/230010.pdf}

Criminal Justice Bill. (2023b, November 14). UK Parliament. \url{https://bills.parliament.uk/bills/3511}

Department of Defense, CISA. (2023, September 12). \textit{Contextualizing Deepfake Threats to Organizations.} \url{https://media.defense.gov/2023/Sep/12/2003298925/-1/-1/0/CSI-DEEPFAKE-THREATS.PDF}

Department of Homeland Security. (2021, September 14). \textit{Increasing Threat of DeepFake Identities.} \url{https://www.dhs.gov/sites/default/files/publications/increasing_threats_of_deepfake_identities_0.pdf}

Doermann, D. (2023, November 8). Testimony before the United States House Committee on Oversight and Accountability, Subcommittee on Cybersecurity, Information Technology, and Government Innovation on 'Recent advances in the creation and distribution of computer-generated images and voice cloning'. \url{https://oversight.house.gov/wp-content/uploads/2023/11/Doermann-Statement41.pdf}

Economic Crime and Corporate Transparency Act 2023, c. 56. \url{https://www.legislation.gov.uk/ukpga/2023/56/enacted}

Egan, J., \& Heim, L. (2023). \textit{Oversight for Frontier AI through a Know-Your-Customer Scheme for Compute Providers.} \url{https://doi.org/10.48550/arXiv.2310.13625}

Fenimore Harper. (2024). \textit{Over 100 deep-faked rishi sunak ads found on meta’s advertising platform.} \url{https://static1.squarespace.com/static/64428d8d19bf554e51471181/t/65a03259148cc61aa541372a/1704997472127/FENIMORE+HARPER+REPORT_+DEEP-FAKED+POLITICAL+ADS+V2.pdf}

Finger, L. (2022, September 8). \textit{Overview Of How To Create Deepfakes - It's Scarily Simple.} Forbes. \url{https://www.forbes.com/sites/lutzfinger/2022/09/08/overview-of-how-to-create-deepfakesits-scarily-simple/}

Fraud Act 2006, c. 35. \url{https://www.legislation.gov.uk/ukpga/2006/35/contents}

Gregory, S. (2023, November 8). Testimony before the United States House Committee on Oversight and Accountability, Subcommittee on Cybersecurity, Information Technology, and Government Innovation on ‘Advances in Deepfake Technology’. \url{https://oversight.house.gov/wp-content/uploads/2023/11/Sam-Gregory-House-Oversight-Committee-Advances-in-Deepfake-Technology-November-2023.pdf}

Home Security Heroes. (2023). \textit{2023 State Of Deepfakes: Realities, Threats, And Impact.} \url{https://www.homesecurityheroes.com/state-of-deepfakes/#key-findings}

H.R.3106 - 118th Congress. (2023-2024). \textit{Preventing Deepfakes of Intimate Images Act.} (2023, May 5). \url{https://www.congress.gov/bill/118th-congress/house-bill/3106}

H.R.5586 - 118th Congress. (2023-2024). \textit{DEEPFAKES Accountability Act} (2023, September 20). \url{https://www.congress.gov/bill/118th-congress/house-bill/5586/text}

H.R.6943 - 118th Congress. (2023-2024). \textit{No AI FRAUD Act} (2024, January 10). \url{https://www.congress.gov/bill/118th-congress/house-bill/6943/text}

\textit{Hustler Magazine, Inc. v. Falwell.} (1988). 485 U.S. 46. \url{https://supreme.justia.com/cases/federal/us/485/46/}

Internet Watch Foundation. (2023, October 1). \textit{How AI is being abused to create child sexual abuse imagery. }\url{https://www.iwf.org.uk/media/q4zll2ya/iwf-ai-csam-report_public-oct23v1.pdf}

Law Commission, \textit{Intimate Image Abuse Consultation Paper} (Law Com No 253, 2021, February 26th) \url{https://cloud-platform-e218f50a4812967ba1215eaecede923f.s3.amazonaws.com/uploads/sites/30/2021/02/Intimate-image-abuse-consultation-paper.pdf}

Law Commission, \textit{Intimate image abuse: a final report} (HC 326, Law Com No 407, 2022, July, 06) \url{https://s3-eu-west-2.amazonaws.com/cloud-platform-e218f50a4812967ba1215eaecede923f/uploads/sites/30/2022/07/Intimate-Image-Abuse-Report.pdf}

Ministry of Justice. (2022, November 25). \textit{New laws to better protect victims from abuse of intimate images.} \url{https://www.gov.uk/government/news/new-laws-to-better-protect-victims-from-abuse-of-intimate-images}

Miotti, A., \& Wasil, A. (2023a). Taking control: Policies to address extinction risks from advanced AI. \url{https://doi.org/10.48550/arXiv.2310.20563}

Miotti, A., \& Wasil, A. (2023b). An international treaty to implement a global compute cap for advanced artificial intelligence. \url{https://doi.org/10.48550/arXiv.2311.10748}

Multiplatform.AI. (2023, December). \textit{OctoML’s Termination of Partnership with Civitai Over AI-Generated CSAM Concerns} \url{https://medium.com/@multiplatform.ai/octomls-termination-of-partnership-with-civitai-over-ai-generated-csam-concerns-e0921fe71f0d}

NBC News. (2023, November 23). \textit{Little recourse for teens girls victimized by AI 'deepfake' nudes.} Chan, M \& Tenbarge, K. \url{https://www.nbcnews.com/news/us-news/little-recourse-teens-girls-victimized-ai-deepfake-nudes-rcna126399}

NBC News. (2024a, January 16). \textit{Teen deepfake victim pushes for federal law targeting AI-generated explicit content.} Tenbarge, K. \url{https://www.nbcnews.com/tech/tech-news/deepfake-law-ai-new-jersey-high-school-teen-image-porn-rcna133706}

NBC News. (2024b, January 22). \textit{Fake Joe Biden robocall tells New Hampshire Democrats not to vote Tuesday.} \url{https://www.nbcnews.com/politics/2024-election/fake-joe-biden-robocall-tells-new-hampshire-democrats-not-vote-tuesday-rcna134984}

\textit{New York Times Co. v. Sullivan.} (1964). 376 U.S. 254. \url{https://supreme.justia.com/cases/federal/us/376/254/}

N.J. Legis. S3707. 2022-2023 (2023, March 9). \textit{Prohibits deepfake pornography and imposes criminal and civil penalties for non-consensual disclosure.} \url{https://www.njleg.state.nj.us/bill-search/2022/S3707}

NO AI FRAUD Act [Draft bill]. (2024, January 9). Reps. María Elvira Salazar \& Madeleine Dean \url{https://files.constantcontact.com/1849eea4801/695cfd71-1d24-4146-a453-3dab7d49babd.pdf}

Onfido. (2024). \textit{Identity Fraud Insights Report 2024.} \url{https://onfido.com/landing/identity-fraud-report/}

Online Safety Act 2023, c. 50. \url{https://www.legislation.gov.uk/ukpga/2023/50/enacted}

Protection of Children Act 1978, c. 37. \url{https://www.legislation.gov.uk/ukpga/1978/37}

Regula. (2023). \textit{Identity Verification Investment: A Study on Its Business Impact.} \url{https://static-content.regulaforensics.com/Landing_pages/the-state-of-identity-verification-2023-report/Identity%20Verification%20Investment-A%20Study%20on%20Its%20Business%20Impact.pdf}

Saberi, M., Sadasivan, VS., Rezaei, K., Kumar, A., Chegini, A., Wang, W., \& Feizi, S. (2023, Sep 29). \textit{Robustness of AI-Image Detectors: Fundamental Limits and Practical Attacks.} \url{https://doi.org/10.48550/arXiv.2310.00076}

Senate Legislative Counsel. (2023, October 11). Draft Copy of EHF23968 GFW [Draft bill]. \url{https://www.coons.senate.gov/imo/media/doc/no_fakes_act_draft_text.pdf}

Smith, B. (2023, September). Testimony before the U.S. Senate Judiciary Committee, Subcommittee on Privacy, Technology, and the Law on “Developing and Deploying AI Responsibly: Elements of an Effective Legislative Framework to Regulate AI” \url{https://www.judiciary.senate.gov/imo/media/doc/2023-09-12_pm_-_testimony_-_smith.pdf}

Stanford Internet Observatory. (2023, December 23). \textit{Identifying and Eliminating CSAM in Generative ML Training Data and Models.} Thiel, D. \url{https://stacks.stanford.edu/file/druid:kh752sm9123/ml_training_data_csam_report-2023-12-23.pdf}

Timmerman, B., Mehta, P., Deb, P., Gallagher, K., Dolan-Gavitt, B., Garg, S., \& Greenstadt, R. (2023). Studying the Online Deepfake Community. \url{https://tsjournal.org/index.php/jots/article/view/126}

Zhang, H., Edelman, B. L., Francati, D., Venturi, D., Ateniese, G., \& Barak, B. (2023a, Nov 7). \textit{Watermarks in the Sand: Impossibility of Strong Watermarking for Generative Models.} \url{https://doi.org/10.48550/arXiv.2311.04378}

Zhang, H., Edelman, B., \& Barak, B. (2023b, November 9). \textit{Watermarking in the sand.} Kempner Institute, Harvard University. \url{https://www.harvard.edu/kempner-institute/2023/11/09/watermarking-in-the-sand/}

\end{document}